\documentclass[twocolumn,showpacs,preprintnumbers,amsmath,amssymb]{revtex4}

\usepackage{graphicx}
\usepackage{dcolumn}
\usepackage{bm}

\begin{document}

\title{The $0$ to $R_1$ cylinder radial coordinate transformation  \\
can not be used to make cylinder layer electromagnetic cloak
}

\author{Jianhua Li}
 \altaffiliation[Also at ]{GL Geophysical Laboratory, USA, glhua@glgeo.com}
\author{  Feng Xie, Lee Xie, Ganquan Xie}%
\affiliation{%
GL Geophysical Laboratory, USA
}%

\hfill\break
\author{Jianhua Li and Ganquan Xie}
\affiliation{
Chinese Dayuling Supercomputational Sciences Center, China \\
Hunan Super Computational Sciences Society, China
}%

\date{June 8, 2017}

\begin{abstract}
In this paper, we discover and proved that the $0$ to $R_1$ cylinder radial coordinate
transformation can not be used to make cylinder layer electromagnetic EM cloak.
Patent of discover,create method and proof in this paper
are reserved by authors in GL Geophysical Laboratory.
\end{abstract}

\maketitle

\section{\label{sec:level1}INTRODUCTION} 
First in the world, in this paper, we discover and prove that the $0$ to $R_1$ cylinder radial $\rho$ transformation method can not be used for making cylinder layer EM cloak. Because in the $0$ to $R_1$ cylinder radial coordinate transformation, the value area of the electric wave field $E_z (\rho ,\phi ,z)$  and magnetic wave field $H_z (\rho ,\phi ,z)$  are invariant. After $0$
to $R_1$ cylinder radial transformation,$\rho _q  = R_1  + Q(\rho )$, $E_z (\rho _q ,\phi ,z) = E_z (\rho ,\phi ,z)$,$H_z (\rho _q ,\phi ,z) = H_z (\rho ,\phi ,z)$
. in the physical inner cylinder boundary , $E_z (R_1 ,\phi ,z) \ne 0$ ,$H_z (R_1 ,\phi ,z) \ne 0$ ,  The cylinder EM wave field will propagation penetrate into the inner cylinder 
$\rho  < R_1 ,\left| z \right| < \infty $ .Therefore, the inner cylinder $\rho  < R_1 ,\left| z \right| < \infty $ . can not be cloaked.  In many published papers on EM cylinder cloak, the $0$ to $R_1$ cylinder radial transformation is wrongfully used for their EM cylinder layer cloak. 

The description of this paper are as follows. Maxwell EM equation in the cylinder coordinate in free space is presented in section 2. $0$ to $R_1$ cylinder radial coordinate transformation can not be used to make EM cloak is proposed in section.
 3. Discussion and conclusion is presented in section 4.

\section {Maxwell EM equation in the cylinder coordinate in free space}
In this section, let the $H_z (\rho ,\phi ,z)$ be magnetic field component in $Z$ direction.
We describe magnetic field equation in the cylinder coordinate free space.

\begin{equation}
\begin{array}{l}
 \frac{\partial }{{\partial \rho }}\rho \frac{{\partial H_z }}{{\partial \rho }} + \frac{{\partial ^2 H_z }}{{\rho \partial \phi ^2 }} \\ 
  + \rho \frac{{\partial ^2 }}{{\partial z^2 }}H_z  + k^2 \rho H_z  = M_s , \\ 
 \end{array}
\end{equation}
  where $\rho$ is radial coordinate, $\phi$ is angular coordinate,  $H_z (\rho ,\phi ,z)$ is the magnetic wave field. $k = \omega \sqrt {\varepsilon \mu } $ is the constant EM wave number in free space , $\omega $  is the angular frequency. $\varepsilon $ is the basic electric permittivity in free space, $\mu$ is the basic magnetic permeability in free space. The magnetic source $M_s$ is induced by the point Delta electric source in 
$\vec e_x  = (1,0,0)$ direction,
\begin{equation}
S(\vec r,\vec r_s ) = \delta (\rho  - \rho _s )\delta (\phi  - \phi _s )\delta (z - z_s )\vec e_x ,
\end{equation}
The incident magnetic wave field $H_{z,i} (\rho ,\phi ,z)$ does satisfy the Maxwell magnetic equation (1) in free space,
\begin{equation}
\begin{array}{l}
 H_{z,i} (\rho ,\phi ,z) \\ 
  =  - \sin \phi \frac{\partial }{{\partial \rho }}g(\rho ,\phi ,z) \\ 
  - \cos \phi \frac{1}{\rho }\frac{\partial }{{\partial \phi }}g(\rho ,\phi ,z), \\ 
 \end{array}
\end{equation}
\begin{equation}
g(\rho ,\phi ,z) =  - \frac{1}{{4\pi }}\frac{{e^{ik_b \sqrt {\left| {\rho  - \rho _s } \right|^2  + (z - z_s )^2 } } }}{{\sqrt {\left| {\rho  - \rho _s } \right|^2  + (z - z_s )^2 } }},
\end{equation}
\begin{equation}
\begin{array}{l}
 \sqrt {\left| {\rho  - \rho _s } \right|^2  + (z - z_s )^2 }  \\ 
  = \sqrt {\rho ^2  - 2\rho \rho _s (\cos \phi  - \phi _s ) + \rho _s ^2  + (z - z_s )^2 } , \\ 
 \end{array}
\end{equation}
\begin{equation}
\begin{array}{l}
 H_{z,i} \left( {\rho ,\phi ,z} \right) \\ 
  = \frac{1}{{4\pi }}\frac{{e^{ik\sqrt {\left| {\rho  - \rho _s } \right|^2  + (z - z_s )^2 } } }}{{\left( {\sqrt {\left| {\rho  - \rho _s } \right|^2  + (z - z_s )^2 } } \right)^2 }} \\ 
 \left( {ik - \frac{1}{{\sqrt {\left| {\rho  - \rho _s } \right|^2  + (z - z_s )^2 } }}} \right) \\ 
 (\rho \sin (\phi ) - \rho _s \sin (\phi _s )), \\ 
 \end{array}
\end{equation}


\section {$0$ to $R_1$ cylinder radial coordinate transformation can not be used to make EM cloak}

\subsection {Homogeneous magnetic wave equation in the cylinder $\rho  \le R_2 $ and $\left| z \right| < \infty$}

For  $R_2$> 0 and $\rho _s  > R_2 $, in the cylinder $\rho  \le R_2 $ and $\left| z \right| < \infty $, the homogeneous magnetic wave equation is
\begin{equation}
\begin{array}{l}
 \frac{\partial }{{\partial \rho }}\rho \frac{{\partial H_z }}{{\partial \rho }} + \frac{{\partial ^2 H_z }}{{\rho \partial \phi ^2 }} \\ 
  + \rho \frac{{\partial ^2 }}{{\partial z^2 }}H_z  + k^2 \rho H_z  = 0, \\ 
 \end{array}
\end{equation}
On the cylinder surface boundary $\rho  = R_2 $, the magnetic wave field $H_z  = H_z (\rho ,\phi ,z)$ and its derivative satisfy the continuous boundary conditions
\begin{equation}
H_z (R_2 ^ -  ,\phi ,z) = H_z (R_2 ^ +  ,\phi ,z),
\end{equation}
\begin{equation}
\frac{\partial }{{\partial \rho }}H_z (R_2 ^ -  ,\phi ,z) = \frac{\partial }{{\partial \rho }}H_z (R_2 ^ +  ,\phi ,z),
\end{equation}
                                                      
\subsection {$0$ to $R_1$ cylinder radial coordinate transformation}
For $R_1>0$, $R_2>R_1$, inside of the cylinder $\rho  \le R_2 $, the $0$ to 
$R_1$ cylinder radial continuous coordinate transformation is
\begin{equation}
\rho _q (r) = R_1  + Q(\rho ),0 \le \rho  \le R_2, 
\end{equation}   
\begin{equation}
\rho _q (0) = R_1 ,Q(0) = 0,
\end{equation}  
\begin{equation}
\rho _q (R_2 ) = R_2, Q(R_2 ) = R_2  - R_1,
\end{equation}  
\begin{equation}
\frac{\partial }{{\partial \rho }}\rho _q (R_2 ) = 1,\frac{\partial }{{\partial \rho }}Q(R_2 ) = 1,
\end{equation} 
The inverse mapping is 
\begin{equation}
\rho  = Q^{ - 1} \left( {\rho _q  - R_1 } \right),R_1  \le \rho  \le R_2, 
\end{equation}  
\subsection{Magnetic equation in the cylinder layer $R_1 \le \rho \le R_2$}
To substitute $0$ to $R_1$ cylinder radial coordinate transformation (10)-(14) into (7), the magnetic wave equation (7) in the free space cylinder $\rho  \le R_2 $ and $\left| z \right| < \infty $ is translated to the following anisotropic magnetic wave equation
in the cylinder layer $R_1 \le \rho \le R_2$,
\begin{equation}
\begin{array}{l}
 \frac{\partial }{{\partial \rho _q }}\frac{{\rho _q }}{{\varepsilon _\phi  }}\frac{{\partial H_z }}{{\partial \rho _q }} + \frac{{\partial ^2 H_z }}{{\varepsilon _\rho  \rho _q \partial \phi ^2 }} \\ 
  + \rho {}_q\mu _z \frac{{\partial ^2 }}{{\partial z^2 }}H_z  + k^2 \rho {}_q\mu _z H_z  = 0, \\ 
 \end{array}
\end{equation}  
     
Where the relative parameters induced by the $0$ to $R_1$  cylinder radial transformation (10-14) are
\begin{equation}
\varepsilon_{\rho} = \mu_\rho   =\frac{\rho }{{\rho _q }}\frac{{d\rho _q }}{{d\rho }},
\end{equation}
\begin{equation}
\varepsilon_\phi = \mu_\phi = \frac{{\rho _q }}{\rho }\frac{{d\rho }}{{d\rho {}_q}},
\end{equation}
\begin{equation}
\varepsilon_z =\mu_z = \frac{\rho }{{\rho _q }}\frac{{d\rho }}{{d\rho _q }},
\end{equation}
on the cylinder surface boundary $\rho = R_2$ the magnetic wave field solution $H_z  = H_z (\rho ,\phi ,z)$,of equation(15) and its derivative satisfy the following continuous boundary conditions
\begin{equation}
H_z (R_2 ^ -  ,\phi ,z) = H_z (R_2 ^ +  ,\phi ,z),
\end{equation}
\begin{equation}
\frac{1}{{\varepsilon _\phi  }}\frac{\partial }{{\partial \rho _q }}H_z (R_2 ^ -  ,\phi ,z) = \frac{\partial }{{\partial \rho }}H_z (R_2 ^ +  ,\phi ,z),
\end{equation}

\subsection {$0$ to $R_1 $ cylinder radial coordinate transformation can not be used to make cylinder layer EM cloak}
\hfill\break \\
${\boldsymbol{Theorem \ 1:}}$ \   Suppose that the magnetic wave $H_z (\rho _q ,\phi ,z)$ does satisfy the magnetic equation (15) with relative anisotropic EM parameters (16-18), and satisfy the field and derivative boundary conditions (19) and (20) that is necessary for no scattering from the cylinder $\rho  \le R_2 $, $\left| z \right| < \infty$,then $H_z \left( {\rho _q ,\phi ,z} \right)$ has the analytic  express as follows:      
\begin{equation}
\begin{array}{l}
 H_z \left( {\rho _q ,\phi ,z} \right) \\ 
  = \frac{1}{{4\pi }}\frac{{e^{ik\sqrt {\left| {Q^{ - 1} \left( {\rho _q  - R_1 } \right) - \rho _s } \right|^2  + (z - z_s )^2 } } }}{{\left( {\sqrt {\left| {Q^{ - 1} \left( {\rho _q  - R_1 } \right) - \rho _s } \right|^2  + (z - z_s )^2 } } \right)^2 }} \\ 
 \left( {ik - \frac{1}{{\sqrt {\left| {Q^{ - 1} \left( {\rho _q  - R_1 } \right) - \rho _s } \right|^2  + (z - z_s )^2 } }}} \right) \\ 
 (Q^{ - 1} \left( {\rho _q  - R_1 } \right)\sin (\phi ) - \rho _s \sin (\phi _s )), \\ 
 \end{array}
\end {equation}
Proof: Substitute the analytic magnetic wave field $H_z \left( {\rho _q ,\phi ,z} \right)$ in (21) into the equation (15), by inverse transformation, $H_z \left( {\rho _q ,\phi ,z} \right)$ is put back to $H_z \left( {\rho ,\phi ,z} \right)$ 
in (6) which is solution of (7). Therefore, $H_z \left( {\rho _q ,\phi ,z} \right)$  does 
satisfy the acoustic equation (15) and does satisfy the no scattering boundary condition (19) and (20), The theorem is proved. In next paper, we will use GLHUA analytical expand method to prove the theorem.

\hfill\break \\

${\boldsymbol{Theorem \ 2:}}$ \ Suppose that the relative electric permittivity and magnetic permeability  are induced by $0$ to $R_1$ cylinder radial transformation in (10)-(14),and incident wave is excited by outside electric point source in (2). The magnetic wave does satisfy the necessary no scattering boundary condition on the outer and inner boundary.The magnetic field $H_z \left( {\rho ,\phi ,z} \right)$  is propagation penetrate into the inner cylinder $\rho  \le R_1 $, $\left| z \right| < \infty$. The inner cylinder $\rho  \le R_1 $, $\left| z \right| < \infty$. can not be cloaked.

Proof:  When the point is on the inner cylinder surface boundary $\rho _q  = R_1 $, and $\left| z \right| < \infty $. Substitute  $\rho _q  = R_1 $ into (21), we have
\begin {equation}
\begin{array}{l}
 H_z \left( {\rho _q ,\phi ,z} \right)|_{\rho _q  = R_1 }  \\ 
  = \left\{ {\frac{1}{{4\pi }}\frac{{e^{ik\sqrt {\left| {Q^{ - 1} \left( {\rho _q  - R_1 } \right) - \rho _s } \right|^2  + (z - z_s )^2 } } }}{{\left( {\sqrt {\left| {Q^{ - 1} \left( {\rho _q  - R_1 } \right) - \rho _s } \right|^2  + (z - z_s )^2 } } \right)^2 }}} \right. \\ 
 \left( {ik - \frac{1}{{\sqrt {\left| {Q^{ - 1} \left( {\rho _q  - R_1 } \right) - \rho _s } \right|^2  + (z - z_s )^2 } }}} \right) \\ 
 (Q^{ - 1} \left( {\rho _q  - R_1 } \right)\sin (\phi ) - \rho _s \sin (\phi _s ))\left. {} \right\}|_{\rho _q  = R_1 }  \\ 
  =  - \frac{1}{{4\pi }}\rho _s \sin (\phi _s )\frac{{e^{ik\sqrt {\left| {\rho _s } \right|^2  + (z - z_s )^2 } } }}{{\left( {\sqrt {\left| {\rho _s } \right|^2  + (z - z_s )^2 } } \right)^2 }} \\ 
 \left( {ik - \frac{1}{{\sqrt {\left| {\rho _s } \right|^2  + (z - z_s )^2 } }}} \right) \\ 
 \end{array}
\end {equation}
\begin {equation}
\frac{1}{{\varepsilon _\phi  }}\frac{\partial }{{\partial \rho _q }}H_z (\rho _q ,\phi ,z)|_{r_q  = R_1 }  = 0.
\end {equation}
For simplicity and without confusion, we remove the subscript $q$ in the inner cylinder, it is naturally, we design $\varepsilon _\rho   = \mu _\rho   = 1$, $\varepsilon _\phi   = \mu _\phi   = 1$, $\varepsilon _z  = \mu _z  = 1$ in the inner cylinder $\rho  \le R_1$ and $\left| z \right| < \infty $ . The magnetic wave $H_z \left( {\rho ,\phi ,z} \right),$ in the inner cylinder does satisfy the following magnetic wave equation
\begin {equation}
\begin{array}{l}
 \frac{\partial }{{\partial \rho }}\rho \frac{{\partial H_z }}{{\partial \rho }} + \frac{{\partial ^2 H_z }}{{\rho \partial \phi ^2 }} \\ 
  + \rho \frac{{\partial ^2 }}{{\partial z^2 }}H_z  + k^2 \rho H_z  = 0, \\ 
 \end{array}
\end {equation}
and does satisfy the continuous boundary condition on the cylinder surface boundary $\rho  = R_1 $, and $\left| z \right| < \infty $.
\begin {equation}
\begin{array}{l}
 H_z \left( {\rho ,\phi ,z} \right)|_{\rho  = R_1 }  \\ 
  =  - \frac{1}{{4\pi }}\rho _s \sin (\phi _s )\frac{{e^{ik\sqrt {\left| {\rho _s } \right|^2  + (z - z_s )^2 } } }}{{\left( {\sqrt {\left| {\rho _s } \right|^2  + (z - z_s )^2 } } \right)^2 }} \\ 
 \left( {ik - \frac{1}{{\sqrt {\left| {\rho _s } \right|^2  + (z - z_s )^2 } }}} \right), \\ 
 \end{array}
\end {equation}

\begin {equation}
\rho \frac{\partial }{{\partial \rho }}H_z (\rho ,\phi ,z)|_{\rho  = R_1 }  = 0,
\end {equation}
The continuous boundary conditions (25) and (26) are necessary for no scattering
from the inner cylinder $\rho  \le R_1 $.$\left| z \right| < \infty$
The magnetic wave $H_z \left( {\rho ,\phi ,z} \right)$ is cylinder symmetry function only depend on the radial $\rho $ and $z$,   

\begin {equation}
H_z \left( {\rho ,\phi ,z} \right) = H_z \left( {\rho ,z} \right),
\end {equation}                                    
The $H_z \left( {\rho ,z} \right)$ does satisfy the following ordinary $0$ order Bessel equation 
\begin {equation}
\frac{\partial }{{\partial \rho }}\rho \frac{{\partial H_z }}{{\partial \rho }} + \rho \frac{{\partial ^2 }}{{\partial z^2 }}H_z  + k^2 \rho H_z  = 0,
\end {equation}      
and the following boundary conditions on the $\rho =R_1$, that is necessary no scattering from inner cylinder $\rho  \le R_1 $ .$\left| z \right| < \infty $
\begin {equation}
\begin{array}{l}
 H_z \left( {R_1 ,z} \right) \\ 
  =  - \frac{1}{{4\pi }}\rho _s \sin (\phi _s )\frac{{e^{ik\sqrt {\left| {\rho _s } \right|^2  + (z - z_s )^2 } } }}{{\left( {\sqrt {\left| {\rho _s } \right|^2  + (z - z_s )^2 } } \right)^2 }} \\ 
 \left( {ik - \frac{1}{{\sqrt {\left| {\rho _s } \right|^2  + (z - z_s )^2 } }}} \right) \\ 
 \end{array}
\end {equation}
                                                           
And
\begin {equation}
\frac{\partial }{{\partial \rho }}H_z (\rho ,z)|_{\rho  = R_1 }  = 0,
\end {equation}
                      
We expand the boundary value of the magnetic wave, $H_z \left( {R_1 ,z} \right)$ in (29) as 
\begin {equation}
H_z \left( {R_1 ,z} \right) = \int_0^\infty  {h_g (R_1 ,k_z )} \cos (k_z z)dk_z ,
\end {equation}
We expand the magnetic wave solution of (28),$H_z \left( {\rho ,z} \right)$  as
\begin {equation}
H_z \left( {\rho ,z} \right) = \int_0^\infty  {h(\rho ,k_z )} \cos (k_z z)dk_z ,
\end {equation}
For any $k_z \ge 0$, 
$h(\rho ,k_z )$ does satisfy the following Bessel equation 
\begin {equation}
\frac{\partial }{{\partial \rho }}\rho \frac{{\partial h}}{{\partial \rho }} - \rho k_z^2 h + k^2 \rho h = 0,
\end {equation}
\begin {equation}
h\left( {R_1 ,k_z } \right) = h_g \left( {R_1 ,k_z } \right),
\end {equation}
                         
\begin {equation}
\frac{\partial }{{\partial \rho }}h(\rho ,k_z )|_{\rho  = R_1 }  = 0,
\end {equation}
                     
For $ 0\le k_z \le k$, the solution of (33)-(35), $h(\rho ,k_z )$ is
\begin {equation}
\begin{array}{l}
 h(\rho ,k_z ) = \frac{\pi }{2}\sqrt {k^2  - k_z ^2 } R_1 h_g (R_1 ,k_z ) \\ 
 \left( {N'_0 \left( {R_1 \sqrt {k^2  - k_z ^2 } } \right)J_0 \left( {\rho \sqrt {k^2  - k_z ^2 } } \right)} \right. \\ 
 \left. { - J'_0 \left( {R_1 \sqrt {k^2  - k_z ^2 } } \right)N_0 \left( {\rho \sqrt {k^2  - k_z ^2 } } \right)} \right), \\ 
 \end{array}
\end {equation}
For $ k_z \ge k$, the solution of (33)-(35), $h(\rho ,k_z )$  is
\begin {equation}
\begin{array}{l}
 h(\rho ,k_z ) =  - \sqrt {k_z ^2  - k^2 } R_1 h_g (R_1 ,k_z ) \\ 
 \left( K \right.'_0 \left( {R_1 \sqrt {k^2  - k_z ^2 } } \right)I_0 \left( {\rho \sqrt {k^2  - k_z ^2 } } \right) \\ 
  - I'_0 \left( {R_1 \sqrt {k^2  - k_z ^2 } } \right)K_0 \left( {\rho \sqrt {k^2  - k_z ^2 } } \right)\left. {} \right), \\ 
 \end{array}
\end {equation}
Summary, the magnetic wave solution of (28) ¨C(30) in the inner cylinder,

$H_z (\rho ,z)$ is
\begin {equation}
\begin{array}{l}
 H(\rho ,z) = \frac{\pi }{2}\int_0^k {} \sqrt {k^2  - k_z ^2 } R_1 h_g (R_1 ,k_z ) \\ 
 \left( {N'_0 \left( {R_1 \sqrt {k^2  - k_z ^2 } } \right)J_0 \left( {\rho \sqrt {k^2  - k_z ^2 } } \right)} \right. \\ 
 \left. { - J'_0 \left( {R_1 \sqrt {k^2  - k_z ^2 } } \right)N_0 \left( {\rho \sqrt {k^2  - k_z ^2 } } \right)} \right)dk_z  \\ 
  - \int_k^\infty  {} \sqrt {k_z ^2  - k^2 } R_1 h_g (R_1 ,k_z ) \\ 
 \left( K \right.'_0 \left( {R_1 \sqrt {k^2  - k_z ^2 } } \right)I_0 \left( {\rho \sqrt {k^2  - k_z ^2 } } \right) \\ 
  - I'_0 \left( {R_1 \sqrt {k^2  - k_z ^2 } } \right)K_0 \left( {\rho \sqrt {k^2  - k_z ^2 } } \right)\left. {} \right)dk_z  \\ 
 \end{array}
\end {equation}
                
The theorem 2 is proved. The magnetic field $H_z \left( {\rho ,\phi ,z} \right)$  is propagation penetrate into the inner cylinder $\rho  \le R_1 $, $\left| z \right| < \infty$. The inner cylinder $\rho  \le R_1 $, $\left| z \right| < \infty$. can not be cloaked.

Similarly, The electric wave propoagation penetrate into the inner cylinder, The inner cylinder $\rho  \le R_1 $, $\left| z \right| < \infty$. can not be cloaked.

\section{Discussion and Conclusion}
In many published paper[4], authors proposed $0$ to $R_1$ cylinder radial linear transformation for cylinder layer EM cloak,however the EM wave propagation
penetrate into their inner cylinder $\rho  \le R_1 $, $\left| z \right| < \infty$.  Their inner cylinder can not be cloaked. 
The $0$ to $R_1$
cylinder radial coordinate transformation  
can not be used to make cylinder layer electromagnetic cloak

That necessary no scattering boundary conditions induced that there is unbounded
spherical symmetry magnetic wave $H(\rho ,z) $ propagation in the inner cylinder $\rho < R_1$. 
If the inner magnetic wave $H(\rho ,z) $ meet the some EM scattering object in inner
sphere, for example a fly, the EM scattering wave will propagation
go out outside of the cylinder $\rho \le R_2$. The cylinder $\rho \le R_2$ will be detected.
Also, the linear $0$ to $R_1$ cylinder transformation makes the infinite and exceeding light speed propagation.
In next papers, we will propose a novel transformation [8] for EM cylinder cloak that will
overcome the difficult on the EM wave propagation penetrate into the inner cylinder $\rho < R_1$.
It is totally different from transformation cylinder cloak,our GLHUA double layer cylinder EM incvisible cloak by [1,2,3-5,6,7,8] and GLHUA seismic double cloak and their
exact analytical wave propagation will overcome all difficuties in $0$ to $R_1$ transformation cylinder cloak.
                      
\begin{acknowledgments}
We wish to acknowledge the support of the GL Geophysical Laboratory and thank the GLGEO Laboratory to approve the paper
publication. Authors thank to Professor P. D. Lax for his concern and encouragements  Authors thank to Dr. Michael Oristaglio and Professor You Zhong Guo for their encouragments
\end{acknowledgments}


\end{document}